\documentclass[10pt,conference]{IEEEtran} 
 
\usepackage{color,soul}
\usepackage{tikz}
\usepackage{verbatim}
\usepackage[noadjust]{cite}

\newcommand*\circled[1]{\tikz[baseline=(char.base)]{\textbf{
            \node[shape=circle,draw,inner sep=1pt] (char) {#1};}}}
 
\newcommand{\appr}{\emph{DECREE}}
\newcommand{\RA}{\emph{DECREE-RA}}
\newcommand{\TB}{\emph{DECREE-TB}}
\newcommand{\RP}{\emph{DECREE-RP}}

\newcommand{\paloma}{{PALOMA}}

\definecolor{amethyst}{rgb}{0.6, 0.4, 0.8}

\begin{document}
%
\title{Mobile-App Analysis and Instrumentation Techniques Reimagined with \appr}

\author{\IEEEauthorblockN{Yixue Zhao}
\IEEEauthorblockA{Advisor: Nenad Medvidovic\\
University of Southern California}
}

\maketitle

\begin{abstract}
A large number of mobile-app analysis and instrumentation techniques have emerged in the past decade. However, those techniques' components are difficult to extract and reuse outside their original tools, their evaluation results are hard to reproduce, and the tools themselves are hard to compare. This paper introduces \appr, an infrastructure intended to guide such techniques to be reproducible, practical, reusable, and easy to adopt in practice. \appr ~allows researchers and developers to easily discover existing solutions to their needs, enables unbiased and reproducible evaluation, and supports  easy construction and execution of replication studies. The paper describes \appr's three modules and its potential to fundamentally alter how research is conducted in this area.
\end{abstract}

\section{introduction}
\label{sec:intro}

Current mobile computing research has focused extensively on three threads: \circled{1} \emph{static analysis} techniques that analyze the apps' implementation artifacts statically to extract information of interest (e.g., security vulnerabilities~\cite{felix18fse,lee2017sealant}); \circled{2} \emph{instrumentation techniques} that improve targeted aspects (e.g., performance~\cite{paloma_icse, liu2014characterizing,zhao2017toward}) of an app by directly modifying the app's implementation; and \circled{3} \emph{auxiliary techniques} that analyze external information associated with mobile apps to learn useful lessons (e.g., our recent work that assessed prefetching and caching opportunities~\cite{zhao2018empirically}).

However, 
there is a pronounced gap, specifically between the emergence of static analysis and instrumentation techniques in research and their adoption in practice for four reasons: 
\circled{1} There is no established communication channel between researchers and app developers, thus the techniques may not meet the exact needs in practice and may violate real-world assumptions. 
\circled{2} Research techniques often have steep learning curves, making them difficult to adopt. 
\circled{3} Research techniques are often evaluated in limited settings, rendering any claims insufficiently convincing for app developers to adopt. \circled{4} Existing techniques are usually designed as one-off solutions, making them hard to reproduce, reuse, or customize.

We have faced this gap in our prior research~\cite{lee2017sealant, paloma_icse,zhao2018empirically,zhao2017toward}. The research community at large is also beginning to recognize this gap and the wasted opportunities it causes~\cite{felix18fse,harmanstart,rosefest,acmartifact}. 
Inspired in part by, but going beyond these early efforts, our proposed work---\appr---aims to transform how research in the mobile arena is conducted in order to produce reusable, practical, and reproducible research that is easier to adopt in practice. 
While the concepts behind \appr~ are independent of the specific technology used to develop mobile apps
, we focus on Android due to its dominant market share.

\appr~ is an infrastructure that provides a comprehensive baseline for \underline{\textit{D}}eveloping, \underline{\textit{E}}valuating, \underline{\textit{C}}omposing, \underline{\textit{R}}eusing, \underline{\textit{E}}volving, and \underline{\textit{E}}xploring research techniques in the mobile computing domain, with three research threads: \circled{1} A microservice-based \emph{reference architecture} for static analysis and instrumentation techniques, intended to be comprehensive in scope but simple enough to adopt and tailor. We will evaluate its \emph{reusability} support and \emph{correctness} by migrating existing techniques, and comparing the migrated and original techniques. \circled{2} A corresponding \emph{testbed} to rigorously evaluate and compare static analysis and instrumentation techniques with standard baselines. We will evaluate its  \emph{correctness} and \emph{effectiveness} by comparing the obtained measurements of the original and migrated techniques.  \circled{3} A cloud-based \emph{open repository} that contains \appr-compatible techniques, allowing both researchers and app developers to easily discover what they need, and enabling unbiased comparison and replication studies of \appr-compatible techniques in an automatic manner.  We will evaluate its \emph{correctness} and \emph{performance} by reproducing the evaluations conducted in the second research thread and comparing their results.

\appr~makes the following contributions: \circled{1} a reference architecture to guide the design of mobile computing techniques, so that they can be readily reused by other researchers and adopted by app developers;  \circled{2} a testbed with standard baselines to allow competing techniques to be evaluated fairly and thoroughly; \circled{3} an open repository to bridge the gap between researchers and developers and allow them to leverage each other's knowledge; \circled{4} reproduced evaluation results of exiting techniques to benefit future research and enable replication studies.

The rest of the paper is structured as follows. Section~\ref{sec:approach} details the three research threads. Section~\ref{sec:preliminary} presents our progress to date and obtained evaluation plan. Section~\ref{sec:related} overviews related work and Section~\ref{sec:contribution} concludes the paper.
\section{proposed approach}
\label{sec:approach}

This section describes \appr's three  research threads. 

\subsection{The DECREE Reference Architecture---DECREE-RA}
\label{sec:appr:ra}


We design {\RA} based on the existing static analysis and instrumentation techniques, and our own experience in the mobile computing domain~\cite{zhao2018empirically, paloma_icse, lee2017sealant, felix18fse, ccfg, jimple,liu2014characterizing}. Our aim is to decompose mobile computing techniques into reusable components at a proper granularity with modular design that can, both, serve as a roadmap for future techniques and improve the reusability of  existing techniques. 

\RA's design is based on the microservice architectural style for three reasons. 
\circled{1} The microservice style helps to decouple potentially complex functionality into lightweight, ``black-box'' microservices, which are easy to understand and adopt. 
\circled{2} Existing analysis and instrumentation techniques tend to comprise clearly separable components, and the microservice style would make it easier to reuse such components across techniques. 
\circled{3} The microservice style allows components (i.e., microservices) to be implemented in different programming languages with different technologies, which suits the heterogeneity of the mobile computing domain.

\begin{figure}
	\centering
	\vspace{-3mm}
		\includegraphics[width=0.49\textwidth]{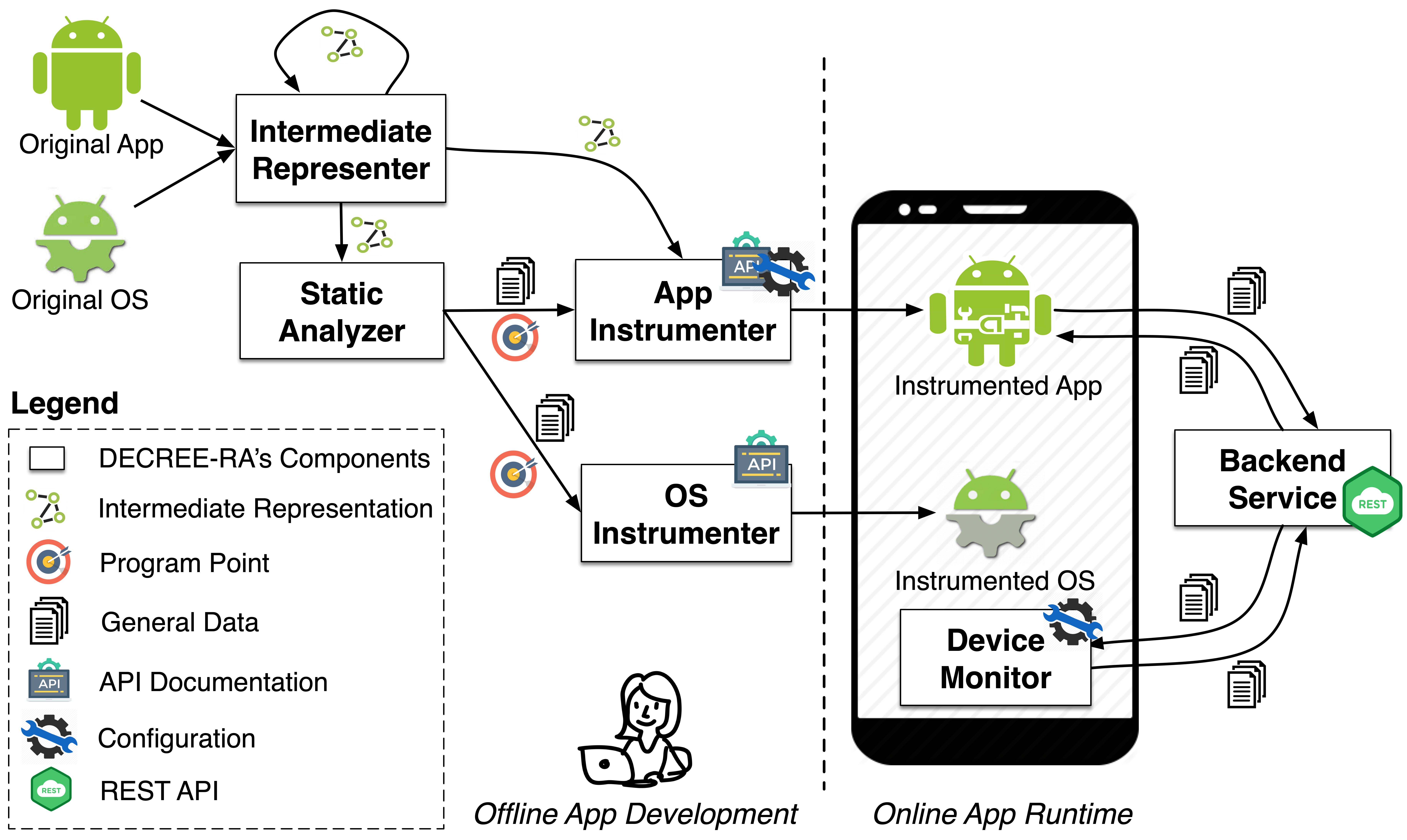}
	\caption{\RA's six reference components and overrall workflow} 
		\vspace{-5mm}
	\label{fig:ra}
\end{figure}

As Fig.~\ref{fig:ra} shows, \RA's reference architecture consists of six components.
An individual static analysis or instrumentation technique can consist of one or more of the reference components. 

\textit{\underline{Intermediate Representer}} takes an app or the OS (e.g., the Android framework), as its input and produces an Intermediate Representation (IR) for \textit{Static Analyzer} to analyze. IR can be used by other \textit{Intermediate Representers} to build new IRs, e.g., a tool-specific IR is usually built on top of foundational IRs, such as the 
control-flow graph (CFG) of an app. 

\textit{\underline{Static Analyzer}} analyzes the IR to extract  information that can be used in other components, such as an app's or OS's program point to be instrumented. For instance, {PerfChecker}~\cite{liu2014characterizing} has a \textit{Static Analyzer} to detect performance bugs. 

\textit{\underline{App Instrumenter}}  transforms the original app, usually based on the information extracted from the \textit{Static Analyzer}. The \textit{App Instrumenter} can be categorized into \textit{Automatic App Instrumenter} or \textit{Manual App  Instrumenter}, e.g., via APIs that leverage annotations, and it usually needs to be configured so that the instrumented app can interact with other specific components at runtime, such as \textit{Backend Service}. 

\textit{\underline{OS Instrumenter}} is similar to \textit{App Instrumenter}, but it instruments the OS (e.g., Android). \textit{OS Instrumenter}s (OSI) can also be categorized as \textit{Automatic} and \textit{Manual} OSIs. For instance, our prior work {SEALANT}~\cite{lee2017sealant}'s \textit{Interceptor} is a Manual OSI that extends the Android framework to block malicious intents at runtime. 

\textit{\underline{Device Monitor}} observes the device-level conditions at app runtime, typically  to balance the quality-of-service (QoS) trade-offs. Similarly to the \textit{App Instrumenter}, it also needs to be configured in order to interact with other components at runtime, such as the \textit{Backend Service}. 

\textit{\underline{Backend Service}} contains the ancillary  functionalities that are triggered at app runtime. It will interact with the instrumented app and the \textit{Device Monitor} via a lightweight protocol, such as REST. The ancillary functionalities are usually triggered by specific information sent from the instrumented app or the \textit{Device Monitor}, such as prefetching HTTP requests aggressively when battery power is  sufficient. 

\subsection{The DECREE Testbed---DECREE-TB}
\label{sec:appr:tb}

\TB~ is a testbed for evaluating both static analysis techniques and instrumentation techniques in a reproducible and unbiased manner. It is intended to support the testing of  techniques that follow \RA's design, and of  apps produced by   instrumentation techniques. 

\subsubsection{Testing of DECREE-Compatible Techniques}
\label{sec:appr:tb:manual}

A  technique can be evaluated at the level of a microservice API with  unit test cases provided by the technique's original developers. 
Each test case is executed in \TB's controlled environment with a built-in monitoring system to record the relevant non-functional properties (NFPs). 

\TB~will store the raw test results of each unit test. These results will be useful for researchers to calculate  coarser-grained evaluation metrics and compare different  techniques. For instance, the \emph{accuracy} of a given technique can be  calculated by the number of relevant \emph{pass} tests. Additionally, \TB's built-in controlled \emph{testing environment} and the NFP monitoring system make it possible to compare different techniques fairly,  under identical conditions.

\subsubsection{Automated Differential Testing}
\label{sec:appr:tb:auto}

\TB~also supports testing an instrumented app,  to verify that the instrumented app's functional behavior is identical to the original app's without unwanted side-effects with desired NFPs (e.g., performance overhead). This is critical but often neglected in the evaluation of existing instrumentation techniques. 
Differential testing has three automated phases: 


\circled{1} In the \textit{differential test generation} phase, the challenge is to efficiently achieve high coverage of the different parts of the apps with confidence. To address  the challenge, we propose a novel \emph{path-sensitive} automatic test generation technique at the granularity of callbacks. Callbacks are the essential representation of user interactions~\cite{ccfg}. For example, the  \texttt{onClick} callback represents a user's click on a GUI widget. Our insight is that the instrumented app should have the same functionalities visible to the users (in addition to the desired NFPs), compared to the original app after each user interaction (i.e., callback). Thus, our test cases aim to cover every possible execution path in the callbacks that contain the difference.  

\circled{2} In the \textit{comparative testing} phase, \TB~will automatically identify the ``checkpoints'' for each test case generated in the previous phase and will run each test case on the original  and instrumented apps to get the results at the checkpoint. 
To render the scope of this research feasible, we will specifically focus on one type of functional and one type of non-functional checkpoints. The functional checkpoint will be inserted at each UI update point, and will be used to verify the instrumented app's correctness. The non-functional checkpoint will be inserted before and after each modified callback, and will be used to verify performance.

\circled{3} The \textit{pair comparison} phase takes pair-wise results generated by \textit{comparative testing}  as input, and compares the results of the original app and the instrumented app for researchers to see if their instrumentation technique works as expected. 


\subsection{The DECREE Repository---DECREE-RP}
\label{sec:appr:rp}

\begin{figure}
	\centering
	\vspace{-3mm}
		\includegraphics[width=0.48\textwidth]{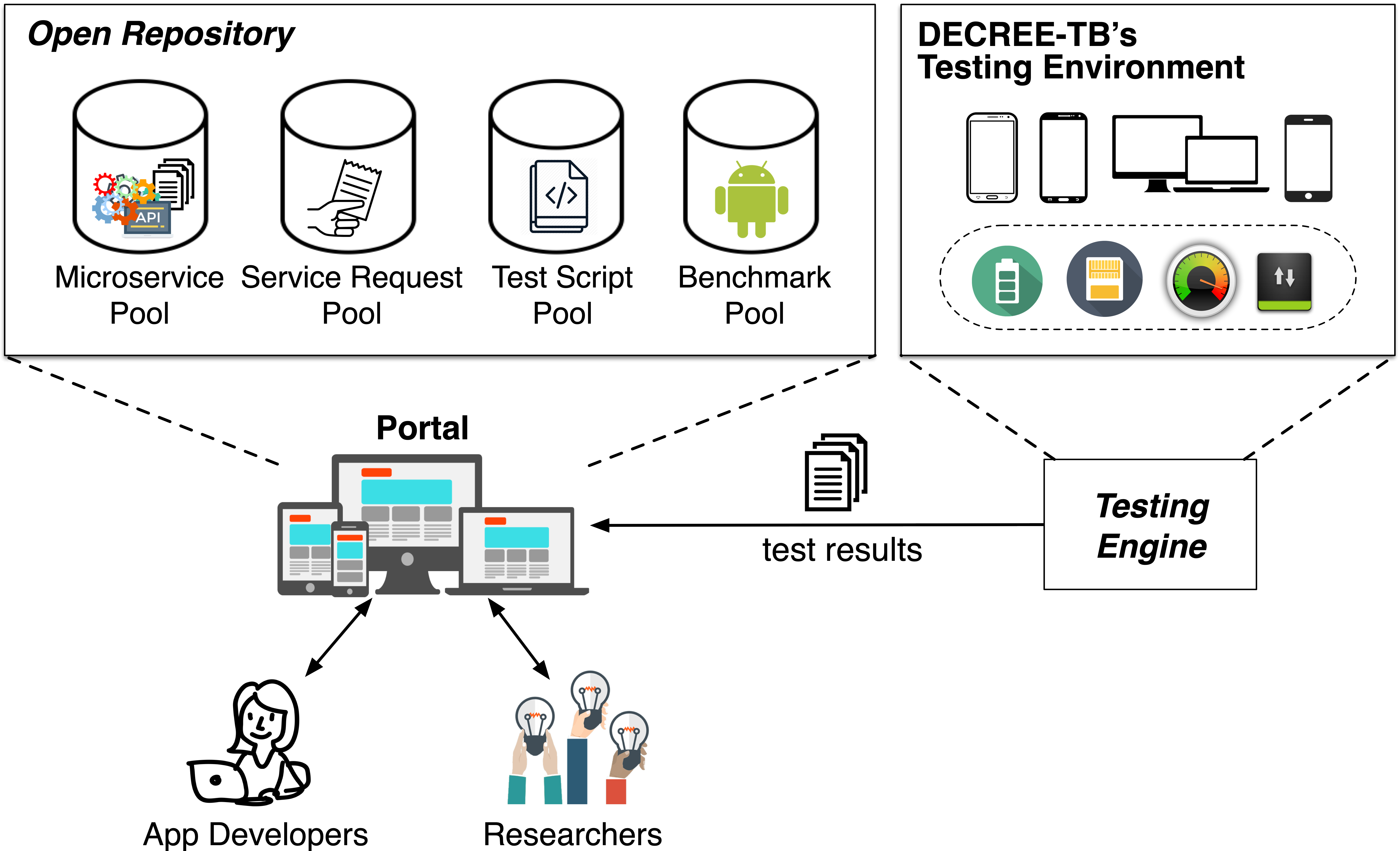}
	\caption{Overview of \RP~ with its open repository and testing engine} 
	\vspace{-6mm}
	\label{fig:rp}
\end{figure}

The goal of \RP~is to improve the availability, reusability, and reproducibility of analysis and instrumentation techniques by providing a cloud-based \emph{open repository} readily accessible to both app developers and researchers, and a built-in \emph{testing engine} integrated from \TB, to enable unbiased, reproducible  evaluation among \appr-compatible techniques in an automatic and customizable manner. 

Fig.~\ref{fig:rp} shows the overview of \RP.
Its \emph{open repository} consists of four databases. \circled{1} \emph{Microservice pool}  contains the micro\-service-based \appr-compatible techniques that are uploaded by  researchers, along with their corresponding API documentation and test results if being evaluated by \TB. 
\circled{2} \emph{Service request pool}  contains the  requests from app  developers on specific capabilities that are needed. 
Developers can  submit test scripts with their service requests to describe their expected results, which can serve as the  ``ground truth'' in the evaluation of research techniques. The test scripts are stored in the \circled{3} \emph{test script pool}, where they can be reused by researchers to determine if their techniques meet the developers' needs, and to compare their techniques with competing techniques under the same baselines. The test script can specify benchmark apps to be tested, which are stored in the \circled{4} \emph{benchmark pool}, where developers or researchers can upload new such  apps to benefit others.

\RP's \emph{testing engine} integrates \TB~and defines a  \emph{test scripting language} to configure the evaluation of \appr-compatible techniques, 
as well as differential testing of  instrumented apps in \TB's controlled testing environment (Section~\ref{sec:appr:tb}). 
The \emph{test scripting language}  specifies \circled{1}~the test cases to be executed; 
\circled{2}~subject apps to be evaluated from the \emph{benchmark pool}; \circled{3}~the testing environment (e.g.,  configurations of the desktop environment for running the \appr-compatible techniques or  versions of the Android device/OS for the apps); and \circled{4}~standard NFP metrics whose monitoring should be enabled  (e.g., execution time). 
With the standard evaluation protocol and a  controlled cloud-based testing environment, \RP's \emph{testing engine} has the potential to alter how the evaluation of research techniques is conducted currently, to enable reproducible and unbiased evaluation and to bypass the often unavoidable time-consuming engineering work (downloading subject apps, setting up controlled testing environments, executing tests and recording their results under varying conditions, etc.).

\section{preliminary results and evaluation plan}
\label{sec:preliminary}

To date, we have developed two mobile computing techniques~\cite{paloma_icse, zhao2018empirically} and designed the \appr~ infrastructure.  We are in the process of migrating our two techniques to \RA, developing \TB~ and  \RP.  The rest of the section describes our progress and evaluation plan in detail.

Our prior work \paloma~\cite{paloma_icse}  is an instrumentation technique, with underlying static analyses, that reduces app latency by prefetching HTTP requests via four major components: (1) \textit{String Analyzer} identifies suitable HTTP requests for prefetching by interpreting their URL values; (2) \textit{Callback Analyzer} detects the program points to issue prefetching requests; (3) \textit{Instrumenter} uses the above information to produce a prefetching-enabled app; (4) at app runtime, the instrumented app triggers \paloma's \textit{Proxy} to issue prefetching requests and cache prefetched responses. 
Following \RA, we are implementing \paloma's  \textit{String Analyzer} and \textit{Callback Analyzer} as two \emph{Static Analyzer} microservices and one reusable Jimple~\cite{jimple} \textit{Intermediate Representer}. \paloma's \textit{Instrumenter} is being implemented as an \textit{App Instrumenter}, while its \textit{Proxy} is being implemented as a \textit{Backend Service} that interacts with the instrumented app at runtime.

Another recent study~\cite{zhao2018empirically} resulted an auxiliary technique with underlying app instrumentation.  
It focused on the prefetching and caching opportunities in mobile apps in order to reduce app latency. It has an \textit{Instrumenter} that instruments the original app to gather  needed information regarding HTTP requests and responses at app runtime, which is used to calculate different statistics for answering research questions, such as ``Are \texttt{Expires} headers trustworthy?''. With \RA, the  \textit{Instrumenter} is being implemented as one \textit{App Instrumenter} microservice, and will reuse the Jimple \textit{Intermediate Representer} microservice developed in \paloma~\cite{paloma_icse}. 

We will  evaluate \RA's support for \emph{reusability} by measuring the common portions in the reimplemented \appr-compatible techniques compared to the original techniques. We will evaluate the \emph{correctness} by verifying that the functionalities of the original techniques remain unchanged in their \appr-compatible counterparts.

We will reuse our prior work~\cite{paloma_icse, zhao2018empirically}  to develop \TB, with the focus on performance (i.e., execution time). Adding other metrics to evaluate (e.g., energy consumption) will be straightforward following the same protocol as performance. \TB's differential testing will leverage our prior experience on program analysis~\cite{paloma_icse} and will be first evaluated on the benchmark apps used originally in our work. We will evaluate its \emph{effectiveness} by comparing the generated test cases with  the ``ground truth'' obtained from the apps' implementations  manually. We will assess the \emph{accuracy} of the applied test cases using the tests reported in our original technique as the baseline. Note that, while \TB's differential testing is motivated by and targeted at instrumented apps, this technique is applicable to any app. Thus, we will further evaluate \TB' \emph{effectiveness} and \emph{accuracy} on a broad cross-section of real Android apps. 


To evaluate \RP, we will  add the re-implemented \appr-compatible techniques, benchmark apps, and test scripts to \RP's \emph{open repository}. Then we will reproduce the same tests conducted in \TB, but this time with the help of test scripts supported by \RP's \emph{testing engine} in order to evaluate its \emph{correctness} and \emph{performance}. 
All results will be recorded and made public throughout, to benefit future  reproducibility studies. 

\vspace{-2mm}
\section{related work}
\label{sec:related}

To the best of our knowledge, we are the first  to propose a comprehensive infrastructure that provides baselines for mobile computing techniques across their  development lifecycle. 
ReproDroid~\cite{felix18fse} proposes a framework for comparing taint analysis tools and reports a reproducibility study on six existing tools. However, ReproDroid 
is limited to taint analysis techniques (one particular type of static analysis techniques) and does not attempt to provide a way of redesigning and reimplementing those techniques for their future improved adaptation and reuse. 
The remainder of this section focuses on software testing, as it is related to \TB~(Section~\ref{sec:appr:tb}). 

\textit{GUI Testing} is widely adopted in testing mobile apps,  such as model-based testing~\cite{su2017guided}, random testing~\cite{monkey}. These techniques target testing the functionalities of an app with high coverage to identify \emph{bugs}, while \TB~ targets  apps that are \emph{instrumented}, often optimized from the original apps.
Thus, \TB's goal is complementary to existing work: instead of achieving high test coverage, it focuses on whether the instrumentation performs as expected with desired NFPs.

\textit{Regression Testing} is a rich research area that focuses on  changes to a program to ensure the changes do not break previous functionalities. It usually assumes that the previous test cases are known, and aims to prioritize or select from the previous test space~\cite{yoo2012regression}, and to adapt or augment the previous test cases to the new changes~\cite{gomez2016mining}. 
\TB's~proposed differential testing technique can generate test cases automatically, without requiring any previous test cases, which would be challenging to obtain for researchers who are not the developers of the apps to be tested.

\section{Expected Contributions}
\label{sec:contribution}

\appr~takes the first step toward open science in the mobile computing domain, with infrastructure support and comprehensive baselines. It has the potential to fundamentally change how app analysis and instrumentation techniques are developed and to yield reusable, reproducible, practical  techniques that benefit both future research and their adoption. An added advantage is that its microservices will be deployed on the cloud and will not introduce significant overhead on the apps deployed on resource-constrained mobile devices. Researchers will also be able to dynamically update their microservices without requiring modifications to the app code. In addition, \appr's test cases have the potential to serve as baselines for comparing different techniques in the same domain (e.g., app optimization for energy efficiency). Once the microservices are adopted by developers, the underlying research techniques will be “organically” evaluated in the real world with real users, providing further insights and incentives for researchers to improve their techniques.
\\


\vspace{-4mm}
\bibliographystyle{IEEEtran}
\bibliography{ICSE2019_DS.bib} 

\end{document}